\documentstyle[11pt,aaspp4]{article}
\def \etal      {et al.}

\begin{document}

\title{Time Resolved Ultraviolet Spectroscopy of DQ Herculis: Eclipses
and Pulsations\footnote{\ Based
on observations with the NASA/ESA Hubble Space Telescope obtained at
the Space Telescope Science Institute, which is operated by the
Association of Universities for Research in Astronomy, Inc., under
NASA contract NAS5-26555.}}

\author{Andrew D. Silber, Scott F. Anderson, Bruce Margon}
\affil{Astronomy Dept., University of Washington, Box 351580, Seattle, WA 98195-1580, USA; \\ 
silber@astro.washington.edu,margon@astro.washington.edu,
anderson@astro.washington.edu}

\and
\author{Ronald A. Downes}
\affil{Space Telescope Science Institute, 3700 San Martin Drive,
Baltimore, MD 21218, USA; \\
downes@stsci.edu}

\begin{abstract} 
The magnetic cataclysmic variable DQ Herculis was observed with the
Faint Object Spectrograph onboard the Hubble Space Telescope (HST)
over four consecutive satellite orbits, including the first observation
in the UV of DQ Her through eclipse minimum. Strong emission in N\,V, C\,IV,
Si\,IV, and He\,II and weak emission in O\,I, S\,III, Ni\,II, N\,IV,
Si\,III, and C\,III was seen.

Time resolved spectroscopy was obtained over 3 orbits with the G160L
grating. At the eclipse minimum, the UV continuum was completely
eclipsed for $< 5\%$ of the orbital period. This implies that the UV
continuum emission region is very compact. By contrast, none of
the emission lines were completely eclipsed, most notably C\,IV which
dropped by only $\sim 75\%$.

DQ Her is known to have intermittent pulsations in the optical and
UV at a period of 71 s, due to the rotation of the accreting white
dwarf. As in our previous observations, the UV continuum and the C\,IV
emission show sporadic 71 s pulsations. The mechanism(s) which
determine the pulsation amplitudes are clearly complex.

The strong emission lines and continuum were auto-correlated during
the first and third HST orbits and  all show a variability timescale of
$\sim 160$ s. The intensities of these lines were also cross-correlated with the
continuum and show a strong correlation with the continuum with a
timelag of $\leq 4$ s. 

The data during the fourth HST orbit were collected with the G190H
grating ($\lambda \lambda$ 1600-2350 with 1.5 \AA\ spectral resolution) in
standard spectroscopy mode (240 s time resolution). These observations resolve 
He\,II $\lambda$ 1640 into a broad (1,400 km s$^{-1}$), asymmetric line.
\end{abstract}


\clearpage
\section{Introduction \& Observations}

Cataclysmic variables (CVs) are semi-detached binary systems
consisting of a red dwarf (the secondary
star) that fills its Roche-lobe  and a white dwarf (the primary) that accretes matter from the
secondary star. DQ Herculis is the prototype of those stars in which
the magnetic field of the white dwarf plays a significant, but not
overwhelming, role (Patterson 1994). The accreting material
forms a disk around the white dwarf, but this disk is disrupted by the
white-dwarf's magnetic field before reaching the surface. In some DQ Her
systems the magnetic field may even disrupt the entire accretion disk
(e.g. Hellier 1991). Unlike the highly magnetic AM Her type CVs in
which the white dwarf rotates in (or nearly in) synchrony with the
orbital period, the white dwarf in DQ Her stars is spinning more
quickly than the orbit, due to accretion torques.

Stable pulsations at a period distinct from the orbital period are the
defining observational property of DQ Her type CVs. These pulsations
originate on the white dwarf at a bright spot where the magnetically
funneled accreting material strikes the white-dwarf surface. The high
temperature and density of the accreting material leads to a high
efficiency of X-ray emission and the highest amplitude of the
pulsations is often in the X-rays. Some of this emission may be
reprocessed elsewhere in the system (e.g. the accretion disk or the
secondary star) which can lead to pulsations in the optical at one or
more of the beat periods between the orbit and spin periods (Warner
1986).

DQ Her itself shows pulsations in the optical broad band emission
(e.g. Walker 1956), the HeII $\lambda$ 4686 line (Chanan, Nelson, \&
Margon 1978; Martell et al. 1995), and the UV continuum and several UV
emission 
lines (Silber et al. 1996; hereafter Paper I).  All of the optical
pulsation amplitudes vary with time in a complicated manner. In this
paper we will show that this is also true of the UV pulsations.

DQ Her itself is only very weakly detected in soft X-rays
(Paper I), and there is no evidence of pulsations in the X-rays. Most
likely this is due to the accretion disk blocking a direct view of the
white dwarf, as would be expected in a system with an inclination this
high ($i = 86.5^\circ$; Horne, Welsh, \& Wade 1993).

DQ Her was observed with the Faint Object Spectrograph (FOS) on the Hubble
Space Telescope over 3 consecutive orbits on 27 April 1995
(post-refurbishment) from 17:09 to 20:52 UT (Table 1). These data were
collected with the $0.9\arcsec$ aperture and the G160L grating in the
``rapid'' mode, with 3.9 s integrations and with spectral
resolution of about 7 \AA. The total integration time was 5,770 s. We
also observed DQ Her for one HST orbit with the G190H grating (from
22:02:02 to 22:42:02) in the standard spectroscopy mode, just after the
G160L data were collected. These data have a spectral resolution of
about 1.5 \AA\ and time resolution of 245 s and were also
collected through the 0.9$\arcsec$ aperture.

Times in this
paper are Heliocentric Julian Date minus 2449800. The
orbital phase is given using the linear ephemeris of Zhang et
al. (1995).  The spin ephemeris used is a linear ephemeris derived
from Zhang et al.'s (1995) cubic ephemeris brought to the epoch of
our observations:
\begin{equation}
T_{max} = 35.3255 + 0.000822 (E).
\end{equation}

\section{Spectrum}

\begin{figure}
\plotone{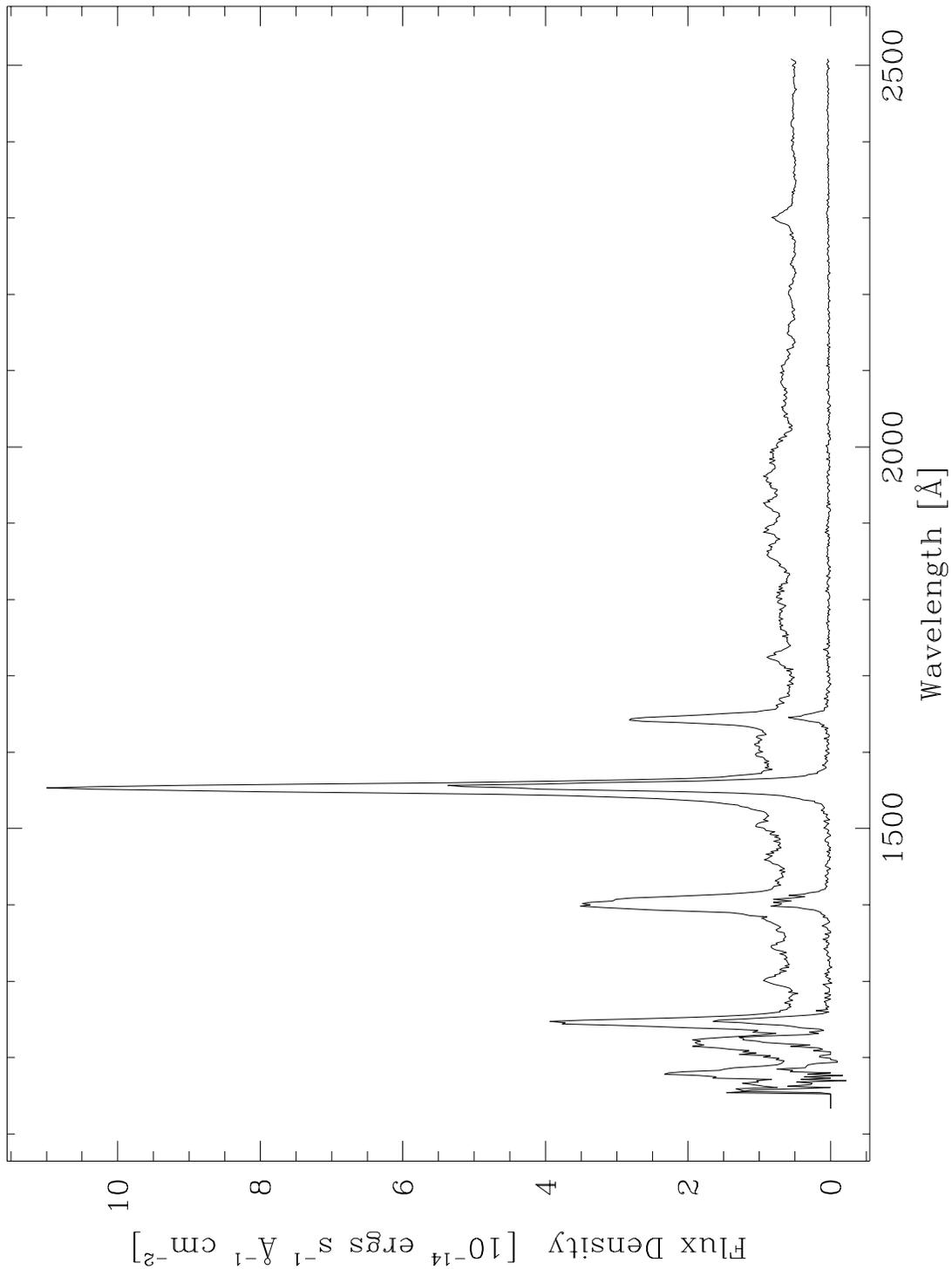}
\figcaption {\label{160spectrum} {\it Upper curve:} The total integrated
spectrum of DQ Her with the G160L grating. Note the strong C\,IV,
Si\,IV, N\,V, and He\,II emission. The strong Ly$\alpha$ emission seen
during the previous visit to DQ Her (Paper I) is not seen, possibly
due to long term variability in the emission from this line or, more
likely, variation in the contamination by the geocoronal
emission. {\it Lower curve:} The spectrum during the eclipse $(35.285
<T<35.297 )$. During the eclipse the continuum emission region is
completely blocked while some line emission is still visible,
especially for C\,IV.}
\end{figure}

The average spectrum of data collected with the G160L grating
(Fig. \ref{160spectrum} and Table 2) shows strong emission of N\,V, C\,IV,
Si\,IV/OIV], and He\,II. Higher-resolution UV spectra taken with the FOS suggest
that the Si\,IV emission dominates the Si\,IV/O\,IV] complex (Eracleous,
private communication). The emission from Ly$\alpha$ is greatly reduced
compared to the previous observation (Paper I) where it was 
concluded that there was significant contamination from geocoronal
emission.

Weak lines of O\,I (1304 \AA) or S\,III] (1298 \AA), Ni\,II (1346),
Ni\,II (1460 \AA), N\,IV (1722 \AA), Si\,III (1892 \AA), and C\,III
(2297 \AA) are also seen. There are many even weaker features,
especially from 1800 to 2000 \AA.  
Except for
the drop in Ly$\alpha$ emission (which is probably explained by
changes in the geocoronal contamination), the spectrum is quite
similar to that seen in Paper I, though the strong emission lines are
more intense in these more recent observations. This brightening cannot
be explained by different orbital sampling because these data do not
include the phase when the hotspot is visible (and the spectrum is
brightest) and do include the eclipse minimum (when the spectrum is
faintest).

The lower spectrum of Fig. \ref{160spectrum} shows the spectrum during the
eclipse. From this it is clear that the UV continuum is completely
eclipsed while the lines are affected to differing degrees. We will
discuss this further in \S \ref{variablity}.

In Paper I we were unable to create a spectrum of reasonable match to
the observations of strong Ly$\alpha$, N\,V, Si\,IV (or OIV]), C\,IV,
and He\,II, emission using the photoionzation code $xstar$ (Kallman \&
McCray 1982; Krolik \& Kallman 1984; Kallman \& Krolik 1986). We have
reconsidered this issue, removing the requirement for strong
Ly$\alpha$. In none of the models are the combination of N\,V, Si\,IV
(or O\,IV]), C\,IV, and He\,II among the 6 strongest lines. Again the
most serious problem is the inability of the models to predict a
strong Si\,IV line. In Paper I one possible explanation given was that
new calculations of the charge transfer process might increase the
production of Si\,IV. Preliminary explorations of this suggest that
the latter process does not significantly increase the production of
Si\,IV 1400 \AA\ (Kallman, pvt. comn).

\begin{figure}
\plotone{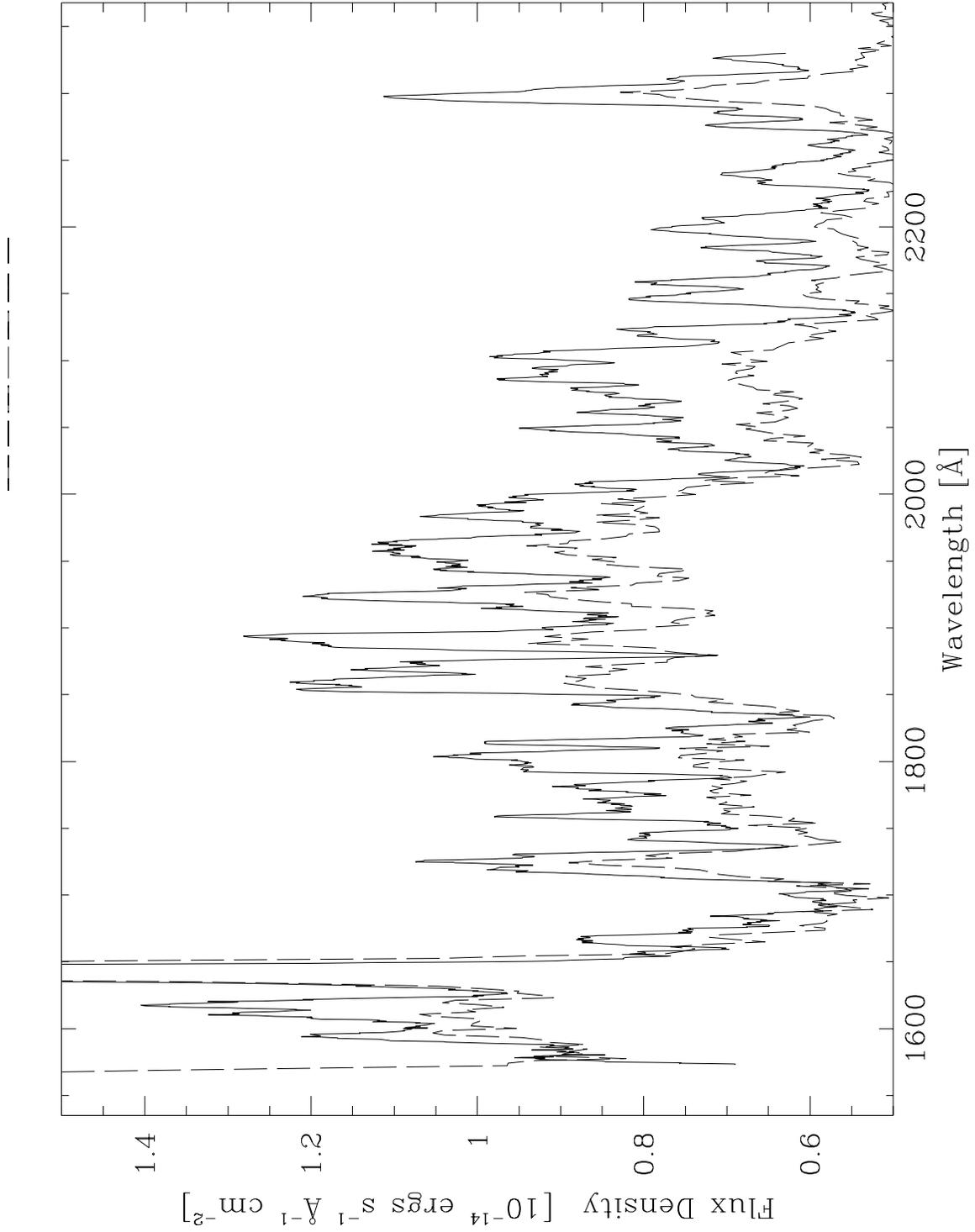}
\figcaption {\label{190spectrum}The spectrum from the G190H (solid)
scaled to
show the weak features including N\,IV (1722 \AA), Si\,III (1892 \AA),
and C\,III (2297 \AA). Numerous weaker features are also apparent and
unidentified, especially in the 1800 to 2000 \AA\ range. We also show
the G160L data (dashed) which has a lower resolution, but shows the
same features.}
\end{figure}

In the G190H spectrum (Fig. \ref{190spectrum}) we see that the weak
lines N\,IV 1722 \AA, Si\,III 1892 \AA, and C\,III 2297 \AA\ are even
more apparent than they were in the low resolution data.  Additionally
there are many weak features in the G190H data that we are unable to
unambiguously identify. These lines are not due to flat-field features in
the detector because they are also visible in the G160L data 
(including the rich complex of features from 1800 to 2000
\AA). One possible source of some of the weak features is Fe\,III in emission,
but a number of the expected lines are seen in absorption or not at all.

\begin{figure}
\plotone{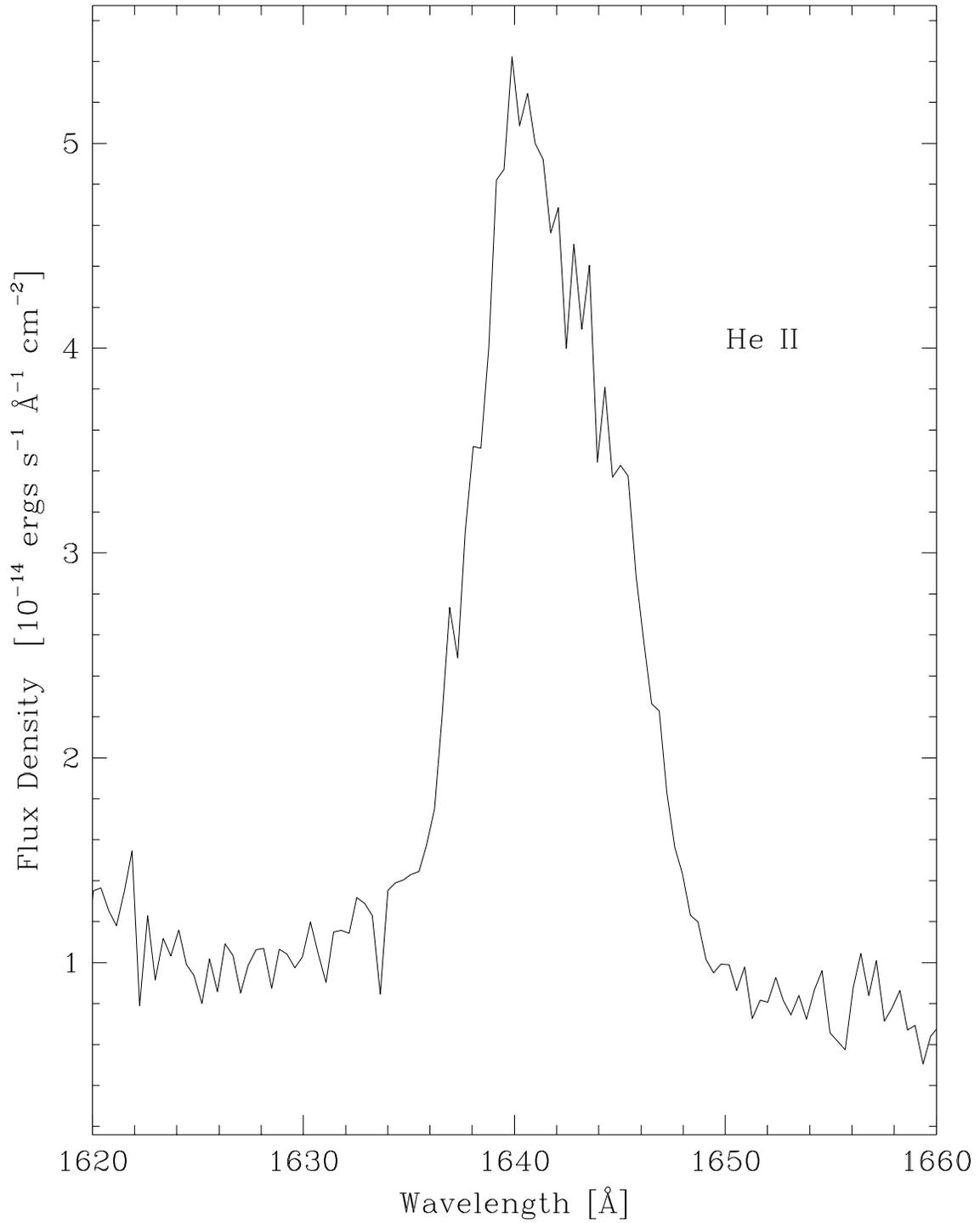}
\figcaption {\label{heii}An expansion around He\,II of the G190H
spectrum. The line is well resolved with a FWHM of 7.7 \AA, compared
to the instrumental resolution of 1.5 \AA. The rise in the blue wing
is much steeper than that in the red.}
\end{figure}

The only strong line included in the G190H data is He\,II 1640 $\lambda$
(Fig. \ref{heii}). This line is clearly resolved with a FWHM of 7.7
\AA\ (1,400 km s$^{-1}$), and an asymmetric profile with a steeper
rise on the blue side. This is similar to the profile of He\,II
$\lambda$ 4686 (Martell et al. 1995). There is no clear variability in
the centroid wavelength or line width of He\,II over the one HST orbit, but the
integrations are too long to show the short timescale variability seen
in the ``rapid'' mode data and the coverage is to short to show
orbital variations.

\section{Time Variability}
\label{variablity}
\begin{figure}
\plotone{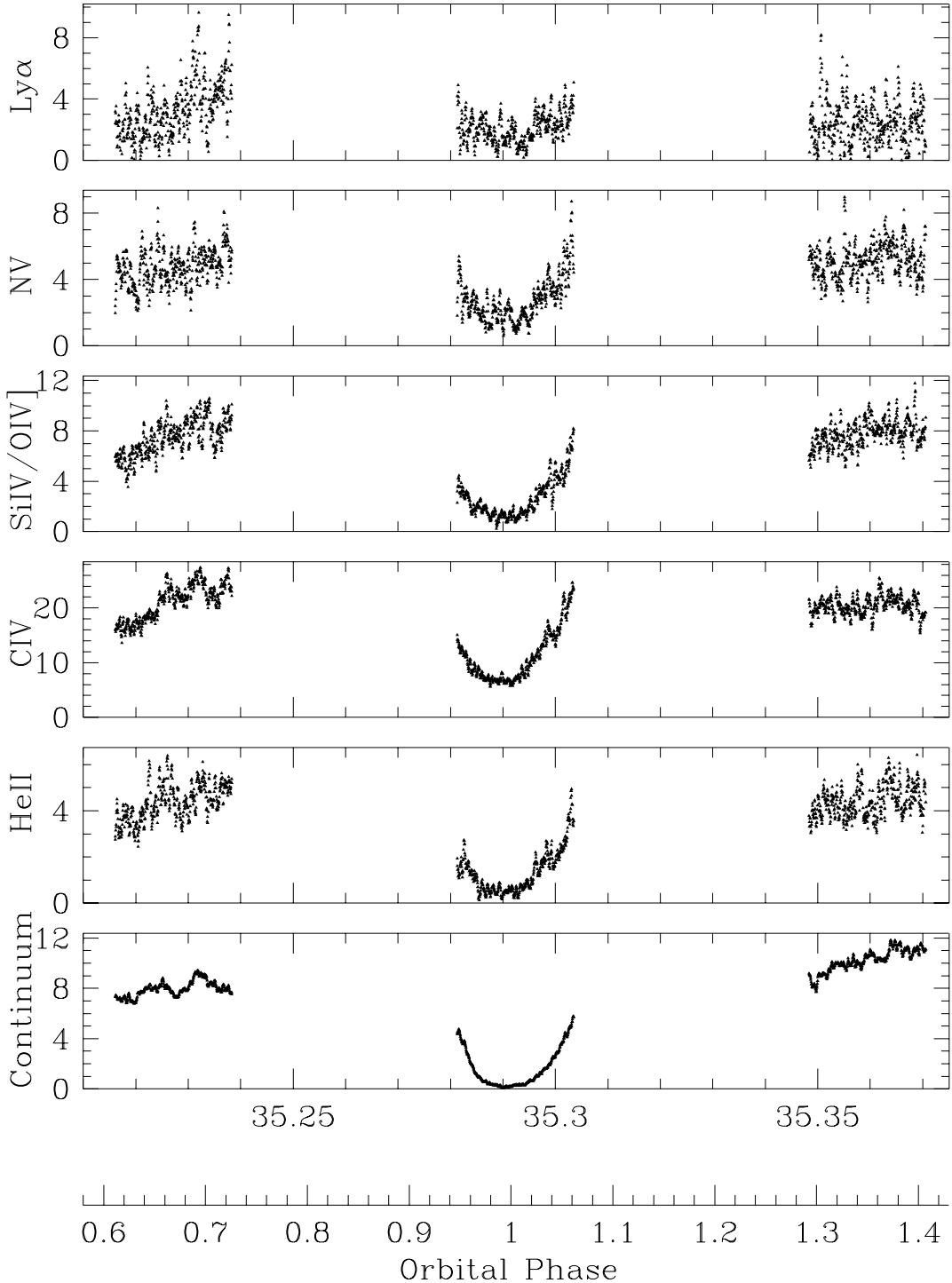}
\figcaption {\label{light}The flux in the 5 strongest lines and the
continuum as a function of time. The continuum includes the flux
from weak lines (e.g. O\,I, C\,III). The eclipse time in the
UV continuum is consistent with the optical eclipse ephemeris. The
line fluxes are significantly, but not completely, diminished during 
the eclipse. There is flickering in both the lines and continuum, which
by inspection can be seen to be correlated, most clearly in the first
HST orbit.}
\end{figure}

In Figure \ref{light} we show the flux of the 5 strongest UV lines and
the continuum from the ``rapid mode'' data. The most striking feature
is the eclipse in the center of the second HST orbit. The UV eclipse
time is consistent with the expected time of the optical eclipse. Two issues
complicate a detailed analysis of the eclipse profile (e.g. eclipse
mapping; Horne 1985): the lack of complete phase coverage; and the
uncertain, but significant, self-occultation by the convex accretion
disk. Nonetheless, we can draw some conclusions from the eclipse
profile.  The symmetry between the ingress and egress argues that the
emission region has a line of symmetry along the line connecting the
two stars. The sloping ingress and egress require that the emission
regions are extended: the continuum egress lasts $>$ 4 \% of the
orbit, which implies a size of $> 1.4 \times 10^{10}$ cm assuming the
orbital parameters of Horne et al. (1993) and that the emission is
centered on the white dwarf. The bulk of the continuum emission region
is eclipsed from 35.286 to 35.295 (about 5\% of the orbital
period). If we ignore the blockage of the white dwarf by the disk, the
predicted duration of the white dwarf eclipse is slightly longer than the
observed UV continuum eclipse, hence the UV continuum eclipse duration
is short, and the size is small.

While the UV continuum is completely eclipsed, the same cannot be said
for the emission lines. The eclipse for He\,II is nearly complete, while
the eclipse for N\,V and Si\,IV are deep, but not complete. Ly$\alpha$
also is partially eclipsed, but due to the low signal-to-noise and
uncertain geocoronal contamination it is unclear what fraction of the
emission from DQ Her is eclipsed.  The most interesting case is C\,IV,
which only drops by about 75\%, showing significant emission even at
the minimum.  Clearly the line emission regions are larger than the
secondary. If we treat the emission region as a uniform sphere
(reasonable for setting lower limits) and adopt the secondary size
given in Horne \etal\ (1993), the C\,IV comes from a region with a
radius of $> 4 \times 10^{10}$ cm. Evidence that C\,IV comes from an
extended region has been seen in other systems (e.g. OY Car; Horne et
al. 1994) as well.

\section{Auto- and Cross-Correlations}

\begin{figure}
\plotone{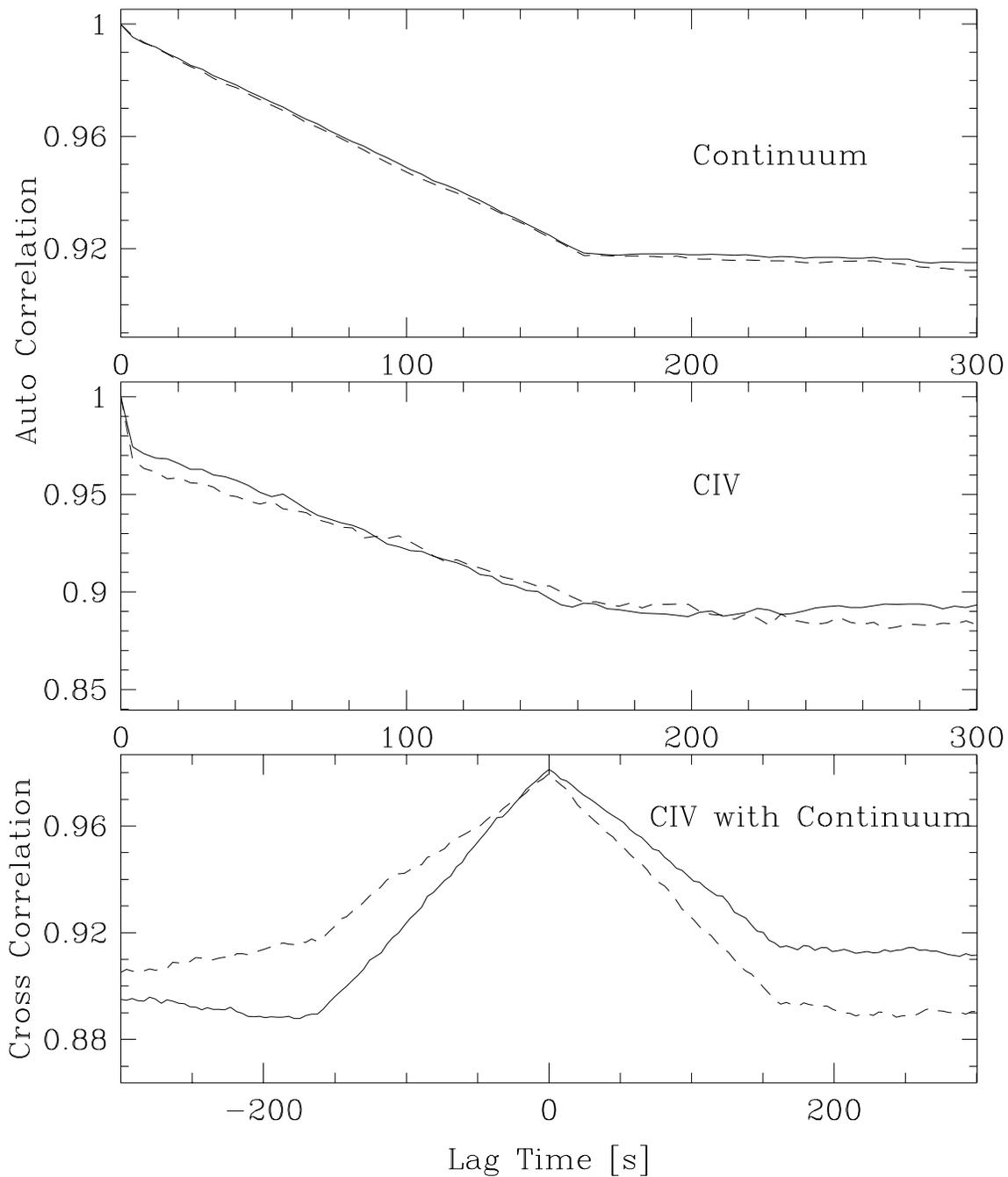}
\figcaption{\label{correlations}The autocorrelation of the continuum
(top) and the C\,IV line (middle) and the cross correlation of the
C\,IV with the continuum (bottom) for HST orbits 1 (solid) and 3
(dashed). The behavior of the other lines is similar to C\,IV line,
although the other lines have inferior signal-to-noise. The
autocorrelation shows evidence of a $\sim$ 2 min timescale. The line
emission is highly correlated with the continuum, while there is no
evidence of a time delay.}
\end{figure}

We performed auto-correlations of the line and continuum emission
during the first and third (non-eclipse) HST orbits separately (Table
3 and Figure \ref{correlations}). These correlations were
normalized such that the value is unity at zero time lag. The results
from the two HST orbits are similar. The auto-correlation of the
continuum slopes from a maximum at zero time lag to a plateau at a lag
time of 160 s. This is similar to what is seen in the optical by
Schoembs \& Rebhan (1989).  The shape of the auto-correlation for the
strong emission lines is similar to that of the continuum, except that
each of the lines has a peak at zero timelag caused by counting noise
which is only correlated at zero timelag. This effect is not important
for the continuum because of the much higher signal-to-noise
ratio. The auto correlation of Ly$\alpha$ is flat, with a spike at
zero timelag. This suggests that the variability of Ly$\alpha$ is
dominated by noise, which would not be correlated at non-zero
timelags, rather than a physical process (e.g. flickering), which is
correlated at small timelags for the other lines.  In Table 3
we list the timescale at which the auto-correlations flatten.

We cross-correlated the emission from the lines with the
continuum (Figure \ref{correlations}). The Ly$\alpha$ emission shows no
correlation with the continuum. All of the other lines are correlated
with the continuum at a high level with a time
delay of $\leq 4$ s. This difference further suggests that the
variability in Ly$\alpha$ is dominated by noise.

\section{The Spin Period}

In Paper I, intermittent pulsations at the white-dwarf spin period (or
half the spin period; Zhang \etal\ 1995) were detected in the
UV continuum, C\,IV, and Ly$\alpha$ emission. We performed a 
search for pulsations and upper limits (Table 4) at the 71-s
period on these newer data as described in Paper I. Amplitude and period errors are estimated via a
Monte Carlo simulation as discussed in Silber \etal\ 1992. The second HST orbit
was split into 3 sections: eclipse ingress ($T<$35.285); eclipse
(35.285 $< T<$ 35.297); and eclipse egress ($T>35.297$).  No pulsation
upper limits are given during HST orbit 2 because of the short
duration of the segments and the lower signal-to-noise.

\begin{figure}
\plotone{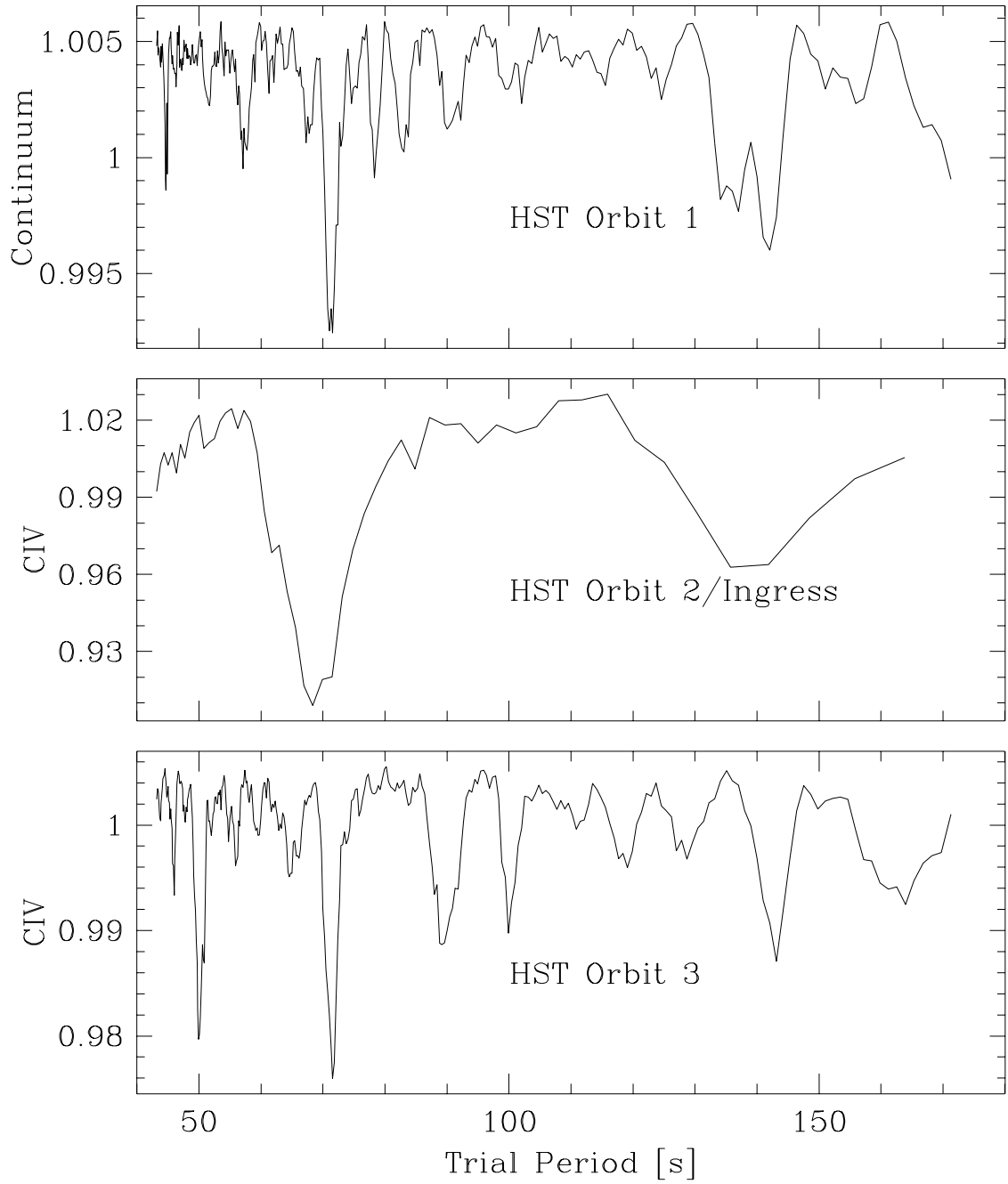}
\figcaption{\label{stel} Pulsations at the
71 s period are clearly detected in two subsets: the
UV continuum during the first HST orbit (top), and C\,IV during the
third (bottom). During the eclipse ingress (middle) there is a weak,
broad signal near the expected period.}
\end{figure}

\begin{figure}
\plotone{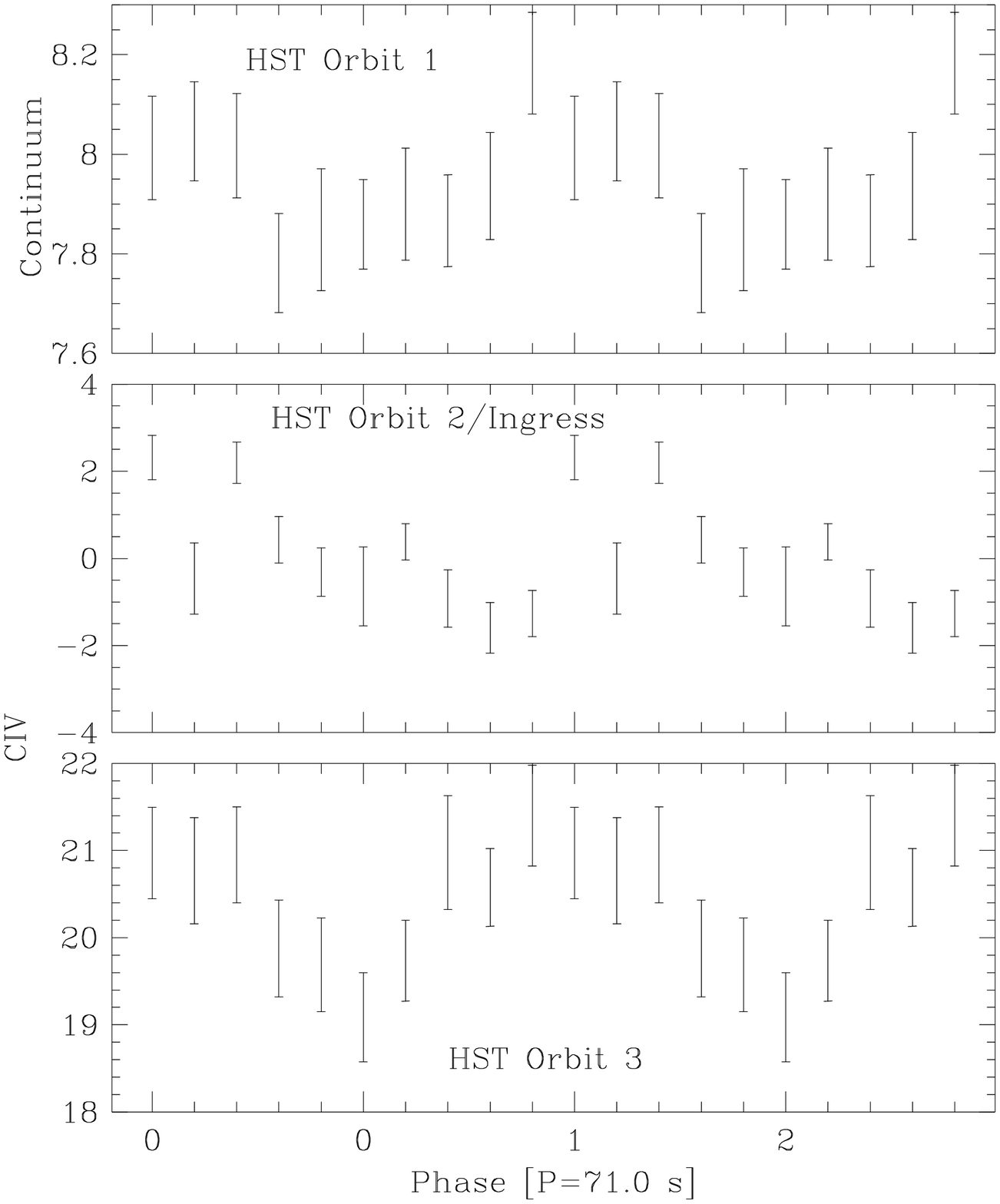}
\figcaption{\label{fold} The continuum during the first HST orbit
 (top) in 10$^{-15}$ ergs s$^{-1}$ cm$^{-2}$ \AA$^{-1}$; C\,IV flux
 during the eclipse ingress after subtracting a linear fit (middle);
 and the C\,IV during the third HST orbit (bottom) in 10$^{-13}$ ergs
 s$^{-1}$ cm$^{-2}$ all folded at a period of 71.0 s.  The
 continuum and C\,IV during the third HST orbit both show clear
 variation that peaks at the same phase as the optical. The eclipse
 ingress pulsation data are less clear and likely not significant.}
\end{figure}

During the first HST orbit the only pulsation detected at the 71-s
period was in the continuum (Fig. \ref{stel}). This pulsation with an
half amplitude of 1.6 $\pm$ 0.6 \% was at 71.3 $\pm$ 0.8 s, consistent with
the expected period of 71.0 s. During the third orbit, C\,IV was seen
to vary at 71.1 $\pm$ 0.5 s with an half amplitude of 4.7 $\pm$ 1.1
\%. C\,IV is also seen to pulse (after removing a linear trend) during
the eclipse ingress at a period of 68.7 $\pm$ 1.3 s and at an
half amplitude of $(1.0 \pm 0.3) \times 10^{-13}$ ergs s$^{-1}$ cm$^{-2}$. It
is unclear whether this pulsation is just due to noise in a short data
set or the rotation of the white dwarf. In Fig. \ref{fold} we see each
of these datasets folded with the ephemeris in Eq. 1.
(1995). We see that the maximum flux from the UV continuum during the first and the C\,IV line during third HST
orbit are maximum at the same phase as the predicted phase of the
optical data. This is contrary to what was seen in Paper I for
C\,IV. As we have no reason to favor one data set over the other, we
infer that the phase of maximum UV flux is yet another parameter which
varies in a complex fashion.

\section{Summary \& Discussion}

DQ Herculis was observed with the HST FOS for three satellite orbits
in April 1995. The spectra were obtained in the ``rapid'' mode with
3.9 s time resolution. The mean spectrum shows strong emission
from N\,V, Si\,IV, C\,IV, and He\,II. These lines are brighter ($\sim
50 \%$) than in our October 1993 observations (Paper I). Unlike our
1993 observations the Ly$\alpha$ emission is weak, suggesting that a
major source of Ly$\alpha$ in the earlier observation was geocoronal,
as we previously inferred. Photoionzation models have difficulty
producing the observed Si\,IV emission.

These data include the first observation in the UV through eclipse
minimum. During the eclipse of the accretion disk, the UV continuum
emission drops to zero; this emitting region must be quite
compact. While the continuum is disappearing, C\,IV, the strongest
line, and N\,V drop by $\sim 75 \%$.  He\,II and Si\,IV are more
strongly eclipsed, but even these lines are still seen at the center
of the eclipse. Clearly the emission lines come from a region
significantly more extended than the continuum. Whatever produces these
lines must be larger than the secondary.

From optical studies (e.g. Chanan et al. 1978; Young \& Schneider
1980) we know that the Balmer lines and He\,II $\lambda$ 4686 are
dominated by disk emission.  The optical continuum is completely
eclipsed, while H$\beta$ is eclipsed by about 75\% (Martell et
al. 1995). H$\gamma$ and He\,II $\lambda$ 4686 are also deeply, though
not completely, eclipsed. Hence the behavior in the UV is similar to
that seen in the optical.

The high level of the cross-correlation between the UV continuum and
the lines argues that the emission between the two is tightly coupled,
even though the eclipse requires that the emission regions are
physically distinct (though possibly overlapping). The similar
timescale of the auto-correlation also argues for a relation between
the variability of the UV lines and continuum. What is unclear is cause
and effect: are the lines brightened by reprocessed UV continuum
emission; are both caused by inhomogeneities in the accretion;
or are both reprocessed X-ray emission? The larger size of the line
emission region suggest that reprocessing is part of the answer. As
the light travel time across the white-dwarf Roche lobe is about 3 s,
higher time resolution is needed to answer this question.

Coherent pulsations at the 71 s spin period are seen in the UV
continuum, but only
during the first HST orbit. C\,IV is seen to pulse
during the third HST orbit. The phasing of both of these pulsations is
consistent with that of the optical pulsations. There is also a
possible detection of pulsations in C\,IV during the eclipse ingress.
Our previous observations of DQ Her also showed intermittent pulsations of the
UV continuum and emission lines. These observations raise the question of
what factor(s) control the complex behavior of the pulsation
amplitudes. Even after four decades of observations in the optical, the
mechanism that controls the amplitude of these pulsations is unknown. What
is clear from our data is that the UV behavior is at least equally complex.

This work was supported by NASA Grant NAG5-1630.

\newpage
\begin{deluxetable}{lccccl}
\tablenum{1}
\tablewidth{0pt}
\tablecaption{Observation Summary}
\tablehead{
\colhead{HST Orbit}  & \colhead{Start [JD-2449800]}  &
\colhead{Duration [s]} &
\colhead{Time Resolution[s]}&
\colhead{Grating} & \colhead{ $\phi_{orbit}$}}
\startdata
1 & 35.22 & 1923 & 3.9 & G160L\tablenotemark{a} & 0.67\\
2 & 35.28 & 1923 & 3.9 & G160L\tablenotemark{a} & 0.01\\
3 & 35.35 & 1923 & 3.9 & G160L\tablenotemark{a} & 0.33\\
4  & 35.42 & 2454 & 245 & G190H\tablenotemark{b} & 0.72
\enddata
\tablenotetext{a}
{$\lambda\lambda 1150-2500$, $\lambda/\Delta \lambda = 260$}
\tablenotetext{b}
{$\lambda\lambda 1600-2350$, $\lambda/\Delta \lambda = 1300$}
\end{deluxetable}

\begin{deluxetable}{lllllll}
\tablenum{2}
\tablewidth{0pt}
\tablecaption{DQ Her Ultraviolet Emission Lines in the Time-Averaged Spectrum}
\tablehead{
\colhead{}  & 
\colhead{Ly$\alpha$} &
\colhead{N\,V}  &
\colhead{Si\,IV} & \colhead{C\,IV}&\colhead{He\,II\tablenotemark{a}}}
\startdata
Flux\tablenotemark{b}	&2.4	&4.2	& 5.7	& 16.4 & 3.0	(3.3)	\\
EW[\AA] 	& 	35	&63	& 86	& 201	& 43 (36)	\\
Wavelength[\AA]	&1218	&1246	& 1402	& 1552	& 1643
(1641)\\
FWHM[\AA]	&20	&12	&19	& 14	& 11 (8)
\enddata
\tablenotetext{a}
{The values in parenthesis are from the fourth HST orbit (G190H data).
All the other data is from the combined G160L spectrum.}
\tablenotetext{b}
{$10^{-13}$ ergs s$^{-1}$ cm$^{-2}$ \AA$^{-1}$}
\end{deluxetable}

\clearpage
\begin{deluxetable}{lll}
\tablecaption{Auto-Correlations Timescale}
\tablenum{3}
\label{auto}
\tablehead{\colhead{Line} & \colhead{Orbit 1 [s]}  & \colhead{Orbit 3 [s]} }
\startdata
Continuum 	&  160 &	160 \\
N\,V		& 160&	140	 \\
Si\,IV		& 160&	200\\
C\,IV		& 150&	180\\
He\,II		& 160& 	180
\enddata
\end{deluxetable}

\begin{deluxetable}{lll}
\tablenum{4}
\label{upper}
\tablewidth{0pt}
\tablecaption{Pulsation Upper Limits}
\tablehead{
\colhead{Line}  & \colhead{HST Orbit 1 [\%]} &
\colhead{HST Orbit 3[\%]}}
\startdata
Ly$\alpha$ & 30	& 25	\\
N\,V	 & 6	& 10	\\
Si\,IV]	 & 8	& 10	\\
C\,IV	 & 4	& N/A	\\
HeII	 & 6	& 9	\\
Continuum & N/A	& 2
\enddata
\end{deluxetable}

\newpage

\clearpage


\begin{references}
Chanan G.A.,  Nelson J.E. \& Margon  B. 1978, ApJ, 226, 963

Hellier, C. 1991, MNRAS, 251, 693

Horne, K. 1985, MNRAS, 213, 129

Horne K., Marsh T.R., Cheng F.H., Hubeny I. \& Lanz T. 1994, ApJ, 426, 294

Horne K., Welsh  W.F. \& Wade  R.A. 1993, ApJ, 406, 229

Kallman, T. R. \& Krolik, J. H.  1986, ApJ, 308, 805

Kallman, T. R. \& McCray, R. 1982, ApJS, 50, 263

Krolik, J. H. \& Kallman, T. R. 1984, ApJ, 286, 366

Martell  P.J., Horne  K., Price  C.M. \& Gomer R.H. 1995, ApJ, 448, 380


Patterson J. 1994, PASP, 106, 481

Schoembs R. \& Rebhan  H. 1989, A\&A, 224, 42

Silber A., Bradt H.V., Ishida M., Ohashi T.  \& Remillard R.A. 1992,
ApJ, 389, 704

Silber, A., Anderson, S. F., Margon, B., \& Downes, R. 1996, ApJ, 462,
428 (Paper I)


Walker M.F. 1956, ApJ, 123, 68 

Warner, B. 1986 MNRAS, 219, 347

Young P. \&  Schneider  D.P. 1980, ApJ,  238, 955

Zhang, E., Robinson, E. L., Stiening, R. F., \& Horne, K.  1995, ApJ, 454, 447

\end{references}
\end{document}